\documentclass{mem}
\usepackage{natbib}\usepackage{txfonts}\usepackage{balance}
\usepackage{graphicx}
\usepackage[a4paper,breaklinks,dvipdfm]{hyperref}
\idline{NN}{1}
\begin{document}

\title{The primordial Li abundance derived from giant stars}

\author{
A. \,Mucciarelli\inst{1}, 
M. \,Salaris\inst{2}
\and P. \, Bonifacio\inst{3}
          }

  %\offprints{Bonifacio}

\institute{
Dipartimento di Astronomia, Universit\'a degli Studi di Bologna, 
via Ranzani 1, 40127, Bologna, Italy\\
\email{alessio.mucciarelli2@unibo.it}
\and
Astrophysics Research Institute, Liverpool John Moores University, 
12 Quays House, Birkenhead, CH41 1LD \\
\email{ms@astro.livjm.ac.uk}
\and
 GEPI, Observatoire de Paris, CNRS, Univ. Paris Diderot, 92125, Meudon Cedex, France\\
\email{piercarlo.bonifacio@obspm.fr}
}

\authorrunning{Mucciarelli et al.}

\titlerunning{Li in lower RGB stars.}

\abstract{

In this contribution we discuss the use of the surface Li abundance in 
lower RGB stars as alternative diagnostic of the primordial Li abundance.
These stars are located in the portion of the RGB after the 
completion of the First Dredge-Up and before the extra-mixing 
episode occurring at the RGB Bump magnitude level.
They are sensitive 
to the total Li content left at the end of the Main Sequence phase 
and are significantly less sensitive to the efficiency of atomic diffusion 
when compared with dwarf stars. 
We analysed lower RGB stars in the Galactic Halo and in the globular clusters 
NGC 6397, NGC 6752 and M4. The final estimates of $\rm{A(Li)}_0$ span a narrow 
range of values (between 2.28 and 2.46 dex), in good agreement with the 
{\sl Spite Plateau} and confirming the discrepancy with 
the values obtained from the standard Big Bang nucleosynthesis calculations.

}
\maketitle{}

\section{Introduction}

The so-called {\sl Lithium problem} is one of the most intriguing and debated 
astrophysical topics, touching on a number of aspects of cosmology and stellar 
evolution. 
Population~II dwarf stars hotter than $\sim$5500 K share the same Li abundance regardless of 
their metallicity and temperature---the so-called {\sl Spite Plateau} 
\citep{spite82}. This evidence --  confirmed by three 
decades of observations -- has been interpreted as the signature of the 
primordial $^{7}$Li abundance produced during the standard Big Bang nucleosynthesis (SBBN).
The {\sl Spite Plateau} turns out to be in the range A(Li)=~2.1--2.4 dex 
(where A(Li) is equal to 12+log(N(Li)/N(H)), depending on the adopted temperature scale.
The recent estimate of the cosmological baryon density obtained with the WMAP 
satellite \citep{larson11} contradicts the classical interpretation 
of the {\sl Spite Plateau}, providing a primordial abundance A(Li)=2.72$\pm$0.06 dex 
\citep{cyburt08}, higher 
than the {\sl Spite Plateau} by at least a factor of 3.

At present, the discrepancy between the WMAP results and the {\sl Spite Plateau} is 
still unsolved and different solutions have been advanced: (1) an inadequacy of the 
SBBN models; %(Cyburt et al.,2010,JACP,10,032 and references therein); 
(2) Population~III stars could have  
burned some of the pristine Li \citep{piau06}; 
(3) Li may be depleted in the photosphere of Population~II dwarf stars by the combined effect of 
atomic diffusion and some competing turbulent mixing \citep{richard05}.

We investigated an alternative route to derive the primordial 
Li abundance, by using the lower Red Giant Branch 
(RGB) stars \citep{mucciarelli12}. These stars are located after the completion of the First 
Dredge-Up (FDU) and before the extra-mixing episode occurring 
at the magnitude level of the RGB Bump.

\section{Why lower RGB stars?}

The surface Li abundance after the FDU is essentially 
a consequence of the dilution due to the increased size 
of the convective envelope after the Main Sequence phase. 
When a star evolves off the Main Sequence, the surface Li abundance 
starts to decrease because the deepening convective envelope 
reaches layers where the Li burning occurs. The depletion ends 
when the convective envelope reaches its maximum depth and thus 
the FDU is complete. 
Thus, the surface Li abundance during the RGB phase remains constant 
until the onset of the extra-mixing episode occuring at the magnitude 
level of the RGB Bump, that leads to a further drop of the surface Li abundance.

From an observational point of view, the existence of a Li {\sl Plateau} 
among the lower RGB stars has been detected in several works, concerning 
field \citep{gratton00,spite06,palacios06} and globular cluster 
\citep{lind08,mucciarelli11} stars. 
However, the use of the RGB Li {\sl Plateau} as empirical diagnostic of the primordial Li abundance 
has been discussed and investigated only recently by \citet{mucciarelli12}.

\subsection{Theoretical advantages of lower RGB stars}

The main advantage of this kind of stars is that the surface Li abundance 
in lower RGB stars depends mainly on the total Li content left in the star after the Main Sequence phase. 
Thus, the observed Li abundance in these stars is affected by the total amount of 
Li (eventually) burned during the Main Sequence (due to the atomic diffusion or 
other mechanisms invoked to moderate the diffusion). 
On the other hand, in the analysis of the dwarf stars we need to have 
a precise description of the physical mechanisms able to transport Li below the convective 
envelope and eventually into the burning zone. 

We investigated the amount of Li depletion $\Delta$(Li) among the 
lower RGB stars 
with respect to the initial Li abundance $\rm{A(Li)}_0$ by using a set 
of stellar evolution models with different metallicities 
(from [Fe/H]=~--3.62 to [Fe/H]=~--1.01 dex)
and with different physical assumptions. For all the metallicities 
the SBBN A(Li) value of 2.72 dex is adopted as initial A(Li). 

\begin{figure}[h]
\resizebox{\hsize}{!}{\includegraphics[width=3.5in]{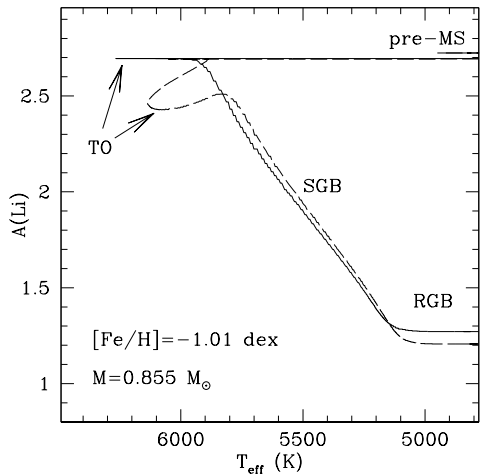}}
\caption{\footnotesize
Evolution of the surface A(Li) as a function of T$_{\rm eff}$ 
for [Fe/H]=~--1.01 dex and 0.855 $M_{\odot}$ stellar models 
with (dashed line) and without (solid line) atomic diffusion.
}
\label{model}
\end{figure}

Fig.~\ref{model} shows the evolution of the surface A(Li) as a function 
of T$_{\rm eff}$ for stellar models
with [Fe/H]=~--1.01 dex and 0.855 $M_{\odot}$,  
with (dashed line) and without (solid line) atomic diffusion. 
The surface Li abundance  at the end of the FDU is only 0.07 dex lower 
when fully efficient atomic diffusion is taken into account.
This is in contrast with models for upper Main Sequence stars, 
where the effect of diffusion on the surface abundances can reach 
several tenths of dex \citep[see e.g. Fig.~3 in][]{mucciarelli11}.

Also, the lower RGB stars are very weakly sensitive to other stellar parameters, 
as the mixing length calibration, the precise stellar age, the initial 
He abundance and the overshooting extension below the Schwarzschild boundary 
of the convective envelope.

Hence, the study of the Li abundance in the lower RGB stars can provide not only an independent 
diagnostic of the primordial Li abundance -- if we consider the discrepancy between the 
Spite Plateau and WMAP results as unsolved -- but also a further strong constraint 
on the efficiency of the additional turbulent mixing, when it is invoked to reconcile Spite Plateau 
with WMAP results.

\section{Primordial Li abundance from lower RGB stars}

\subsection{Halo field stars}
\begin{figure}[h]
\resizebox{\hsize}{!}{\includegraphics[clip=true]{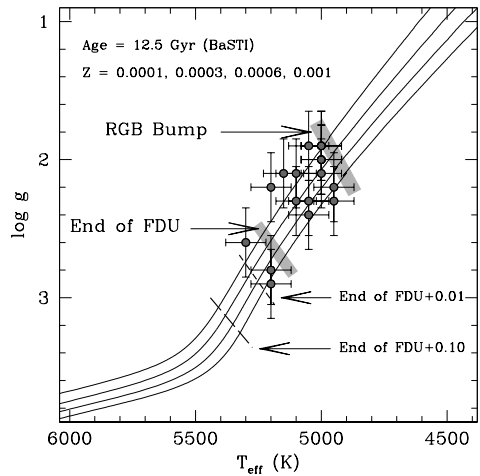}}
\caption{\footnotesize
Position of the targets in the T$_{\rm eff}$--log~g diagram. 
A set of theoretical isochrones at different metallicities 
from the BaSTI database is shown for reference. 
The location of the RGB Bump is marked as a grey shaded region. 
The fainter shaded region marks the boundary beyond which surface
A(Li) stops decreasing. The short-dashed and long-dashed lines 
correspond to A(Li) increased by 0.01 and 0.1 dex, respectively.
}
\label{target}
\end{figure}

We have selected a sample of 17 metal-poor (with [Fe/H] between $\sim$--3.4 and $\sim$--1.4 dex) 
Halo stars located in the lower RGB. 
The position of the targets in the T$_{\rm eff}$--log~g diagram is shown 
in Fig.~\ref{target}.
All the stars are located below the RGB Bump level (thus before the 
onset of the extra-mixing episode), as confirmed also by the measurement 
of the $^{12}C/^{13}C$ ratio (higher than $\sim$15--20).

High resolution (R$>$40000) spectra were retrieved from the ESO
\footnote{http://archive.eso.org/eso/eso$\_$archive$\_$main.html}
 and ELODIE
\footnote{http://atlas.obs-hp.fr/elodie/}
  archives. 
The analysis was performed by employing three different temperature scales, namely 
{\sl (i)}~the photometric scale by \citet{alonso99} (A99), 
{\sl (ii)}~the photometric scale by \citet{ghb09} (GHB09), and 
{\sl (iii)}~the spectroscopic scale based on the excitation equilibrium.
Basically, photometric and spectroscopic scales well agree with each other, regardless 
of the metallicity.

\begin{figure*}[t!]
\resizebox{\hsize}{!}{\includegraphics[clip=true]{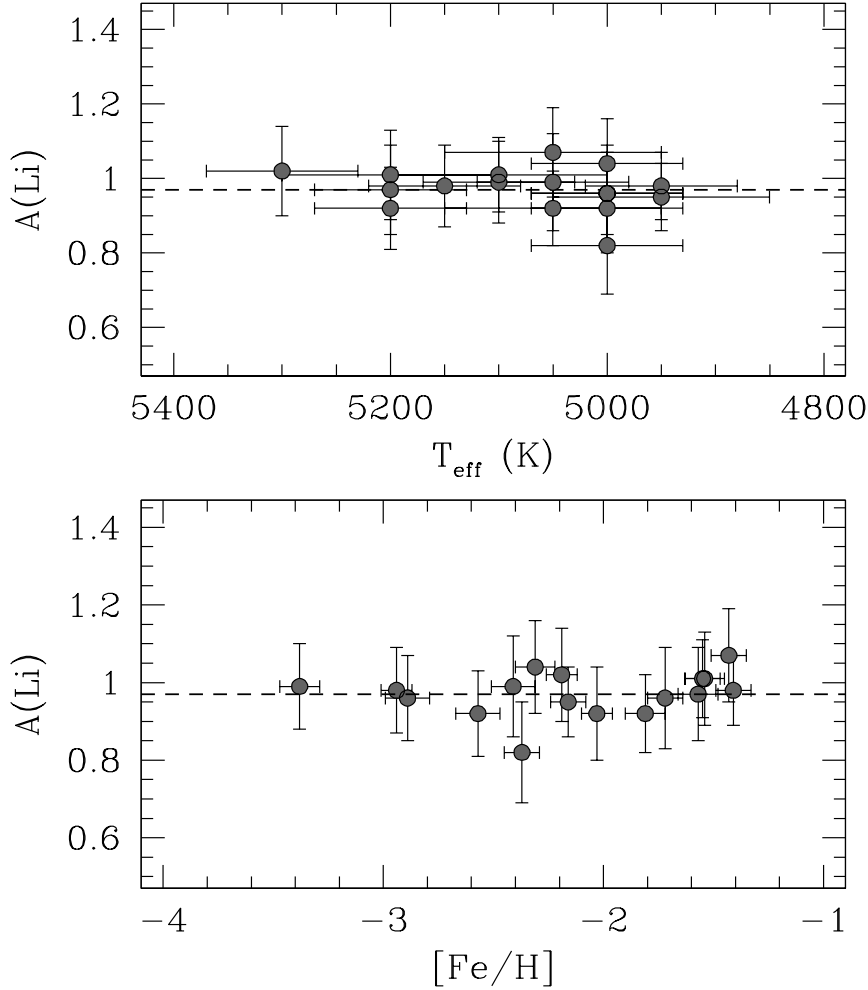}}
\caption{\footnotesize
Behaviour of the Li abundance as a function of the temperature
(upper panel) and of the iron abundance (lower panel). 
The dashed lines represent the average $\rm{A(Li)}_0$ 
of the field stars of the sample.
}
\label{resu}
\end{figure*}

Li abundances have been determined from the Li line at 6707.7 $\mathring{A}$ 
through the comparison with synthetic spectra and applying suitable corrections 
for the departures from LTE.
Fig.~\ref{resu} shows the behaviour of the surface A(Li) as a function of 
the iron content (lower panel) and of the spectroscopic T$_{\rm eff}$ (upper panel). 
The average Li abundance is A(Li)=~0.97 dex ($\sigma$=~0.06 dex), when the 
spectroscopic effective temperatures are adopted, and A(Li)=~0.97 dex ($\sigma$=~0.07 dex) 
and 1.07 dex ($\sigma$=~0.07 dex) by adopting A99 and GHB09 scales, respectively.

Table ~\ref{lit} summarises the estimates of the initial Li abundance $\rm{A(Li)}_0$ 
in our sample of lower RGB stars and considering three sets of stellar evolution 
models (without and with atomic diffusion but without overshooting and 
without diffusion but including overshooting) and the three adopted T$_{\rm eff}$ scales.
The derived $\rm{A(Li)}_0$ span a narrow range of values and our results well match 
the typical values of the {\sl Spite Plateau}.
Models including atomic diffusion (without overshooting) lead to $\rm{A(Li)}_0$ larger by 
0.07 dex with respect to models without diffusion. Models including 
overshooting predict Li abundances larger by only 0.01 dex with respect to 
the standard case. 

\begin{table*}[h]
\caption{Estimates of the initial Li abundance $\rm{A(Li)}_0$ in our sample 
of lower RGB field stars.}
\label{lit}
\begin{center}
\begin{tabular}{lccc}
\hline
\\
Models & $\rm{A(Li)}_0$(A99) &  $\rm{A(Li)}_0$(GHB09)& $\rm{A(Li)}_0$(spect)  \\
\hline
\\
Standard      & 2.28 & 2.39 & 2.30 \\
Diffusion     & 2.35 & 2.46 & 2.37 \\
Overshooting  & 2.29 & 2.40 & 2.31 \\
\hline
\end{tabular}
\end{center}
\end{table*}

\subsection{Globular clusters stars}
In order to strengthen our results, we derived Li abundances in
lower RGB stars in three Galactic globular clusters, namely 
NGC~6397, NGC~6752 and M4. 
%The estimate and interpretation of the initial 
%Li abundance of globular clusters in terms of halo primordial A(Li) 
%should be considered with caution,
%\begin{itemize}
\subsubsection{NGC 6397} --- We have analysed 45 lower RGB stars in NGC~6397, 
retrieved from the dataset of GIRAFFE/FLAMES spectra already discussed by \citet{lind08}.
We derived initial Li abundances $\rm{A(Li)}_0$=~2.33-2.42 dex with the model 
without atomic diffusion (A99 and GHB09, respectively), and 2.40-2.49 dex 
when the model with atomic diffusion is employed.

\subsubsection{NGC 6752} --- We have analysed 21 lower RGB stars retrieved from the UVES archive. 
The models without and with the inclusion of the atomic diffusion provide 
initial Li abundances $\rm{A(Li)}_0$=~2.29-2.35 (GHB09 scale), 2.18-2.24 (A99 scale) and 
2.19-2.25 dex (spectroscopic scale).

\subsubsection{M4} --- The evolution of the surface A(Li) in M4 from the turn-off stars 
to the RGB Bump magnitude level has been discussed in \citet{mucciarelli11}. The derived 
Li abundance in lower RGB stars is A(Li)=~0.92 dex that leads to a primordial Li abundance 
$\rm{A(Li)}_0$=~2.35 dex (without atomic diffusion) and 2.40 dex (with atomic diffusion).
%\end{itemize}
%\section{Lower RGB stars: some advantages}

\section{Conclusions}
In this contribution we have discussed 
the use of the lower RGB stars as alternative route 
to derive the primordial Li abundance 
\citep{mucciarelli12}. 
These stars are sensitive 
to the total Li content left at the end of the Main Sequence phase 
and are significantly less sensitive to the efficiency of atomic diffusion 
(if compared with the dwarf stars). 

The values of $\rm{A(Li)}_0$ inferred from our sample range from 2.28 
(obtained with the A99 T$_{\rm eff}$ scale and without 
the inclusion of atomic diffusion) to 2.46 
(obtained with the GHB09 T$_{\rm eff}$ scale and including atomic diffusion).
When a given T$_{\rm eff}$ scale is adopted, the effect of fully efficient 
atomic diffusion on the derived $\rm{A(Li)}_0$ is by at most 0.07 dex. 
The analysis performed on lower RGB stars in three Galactic globular clusters 
(namely, NGC~6397, NGC~6752 and M4) confirms these values of $\rm{A(Li)}_0$.

These values well agree with those usually obtained by analysing the dwarf stars 
of the {\sl Spite Plateau}. Our results provide an independent estimate of the 
primordial Li abundance, confirming the discrepancy occuring with the 
WMAP+SBBN calculations.

Summarising, the lower RGB stars represent an empirical diagnostic 
of the primordial Li abundance:
\begin{itemize}
\item alternative to the {\sl Spite Plateau}, allowing also estimates of the Li content 
in stellar systems more distant than those usually observed to investigate 
the dwarf stars. Thus, the use of these stars will allow to enlarge the sample 
of field stars and globular clusters to study the primordial $\rm{A(Li)}_0$ abundance within 
the Galaxy but also to assess whether the Li problem exists in 
extragalactic systems \citep[see e.g.][for the study of the initial Li in $\omega$Cen]{monaco10}.
\item complementary to the {\sl Spite Plateau}. In fact, the combination 
of the information arising from the two {\sl Plateau} sets robust constraint for the physical 
processes invoked to resolve the Li 
discrepancy (in particular the efficiency of the turbulent mixing). 
%In fact, a theoretical model of the evolution of the surface Li abundance 
%as a function of the evolutionary stage 
%We stress that a theoretical model of the surface Li abundance 
\end{itemize}

%\begin{acknowledgements}
%Be extremely nice here and acknowledge everybody.
%Forgotten someone?
%\end{acknowledgements}

\bibliographystyle{aa}

\begin{thebibliography}{}
\bibitem[Alonso et al.(1998)]{alonso99} 
Alonso, A., Arribas, S., \& Martinez-Roger- C., 2008, A\&AS, 131, 209

\bibitem[Cyburt et al.(2008)]{cyburt08}
Cyburt, R. H., Fields, B. D., \& Olive, K. A., et al., 2008, JCAP, 11, 12

\bibitem[Gonzalez Hernandez \& Bonifacio(2009)]{ghb09}
Gonzalez Hernandez, J. I., \& Bonifacio, P., 2009, A\&A, 497, 497

\bibitem[Gratton et al.(2000)]{gratton00} 
Gratton, R. G., Sneden, C., Carretta, E., \& Bragaglia, A., 2000, A\&A, 354, 169

\bibitem[Larson et al.(2011)]{larson11}
Larson et al., 2011, ApJS, 192, 16

\bibitem[Lind et al.(2008)]{lind08}
Lind, K., Korn, A. J., Barklem, P. S., \& Grundhal, F., 2008, A\&A, 490, 777L

\bibitem[Monaco et al.(2010)]{monaco10}
Monaco, L., Bonifacio, P., Sbordone, L., Villanova, S., \& Pancino, E., 2010, 
A\&A, 519, L3

\bibitem[Mucciarelli et al.(2011))]{mucciarelli11} 
Mucciarelli, A., Salaris, M., Lovisi, L., Ferraro, F. R., Lanzoni, B., Lucatello, S., 
\& Gratton, R. G., 2011, MNRAS, 412, 81

\bibitem[Mucciarelli et al.(2012))]{mucciarelli12} 
Mucciarelli, A., Salaris, M. \& Bonifacio, P., 2012, MNRAS, 419, 2195

\bibitem[Palacios et al.(2006)]{palacios06} 
Palacios, A., Charbonnel, C., Talon, S., \& Siess, L., 2006, A\&A, 453, 261

\bibitem[Piau et al.(2006)]{piau06}
Piau, L., Beers, T. C., Balsara, D. S., Sivarani, T., Truran, J. W., \& Ferguson, J. W., 2006, ApJ, 653, 300

\bibitem[Richard et al.(2005)]{richard05}
Richard, O., Michaud, G., \& Richer, J., 2005, ApJ, 619, 538

\bibitem[Spite \& Spite(1982)]{spite82}
Spite, M. \& Spite F., 1982, A\&A, 115, 357

\bibitem[Spite et al.(2006)]{spite06} 
Spite, M., et al., 2006, A\&A, 455, 291

\end{thebibliography}

\end{document}